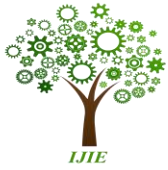

**International journal of innovation in Engineering**
journal homepage: www.ijie.ir

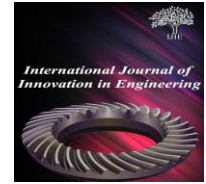

**Research Paper**

# Providing a model for the issue of multi-period ambulance location

Hamed Kazemipoor[*][a], Mohammad Ebrahim Sadeghi[b], Agnieszka Szmelter-Jarosz[c], Mohadese Aghabozorgi[d]

[a] Department of Industrial Engineering, Islamic Azad University, Central Tehran Branch, Iran

[b] Department of Industrial Management, Faculty of Management, University of Tehran, Tehran, Iran

[c] Department of Logistics, Faculty of Economics, University of Gdańsk, Poland

[d] Bachelor of Industrial Engineering, Islamic Azad University, Central Tehran Branch, Iran

**A B S T R A C T**

In this study, two mathematical models have been developed for assigning emergency vehicles, namely ambulances, to geographical areas. The first model, which is based on the assignment problem, the ambulance transfer (moving ambulances) between locations has not been considered. As ambulance transfer can improve system efficiency by decreasing the response time as well as operational cost, we consider this in the second model, which is based on the transportation problem. Both models assume that the demand of all geographical locations must be met. The major contributions of this study are: ambulance transfer between locations, day split into several time slots, and demand distribution of the geographical zone. To the best of our knowledge the first two have not been studied before. These extensions allow us to have a more realistic model of the real-world operation. Although, in previous studies, maximizing coverage has been the main objective of the goal, here, minimizing operating costs is a function of the main objective, because we have assumed that the demand of all geographical areas must be met.

---

[*] Corresponding author:
hkazemipoor@yhoo.com

# 1. Introduction

Optimal location of emergency vehicles, in particular, in big cities is a very strategic and vital decision, as if has a dramatic impact on quality of life and people lives. In the past 30 years, many researchers have worked on this problem and developed different modelling approaches, objectives and criteria, and solution strategies as well as considering real-world and operational constraints. Heuristic and meta-heuristic algorithms have been of great attention for optimally locating ambulances, as well as for other optimization problems. One of the problems in the context of locating emergency vehicles, is locating ambulances to maximize zone coverage as well as maximizing quality of service to patients, in particular under emergency scenarios. As we know, the goal of such services is to minimize the fatality and injuries by giving effective and prompt services to patients in need. On the other hand, there are constraints and limitations imposed by organizational and managerial sides as well as real-world conditions that make the problem challenging. Some of these constraints and challenges might be unknown rate and time of demand, unknown travel times, unseen incidents such as traffic jams, etc. To this extent, optimal location of ambulances such that they will reach patients has dramatic impact on people lives, and therefore, on the community as a whole (Handler and Mirchandani (1979)). The ambulance location problem is to locate ambulances such that the demand for ambulances can be met under pre-set time standards, such that at least a certain fraction of the demand zone is covered. As the need to ambulances and the associated services vary by the time of the day as well different days of the planning horizon, optimal location of ambulances can improve system performance by answering to such a need (Daskin,1995).

We can categorize the ambulance location models into two groups of 'static location problems' and 'dynamic location problems'. Following this, two sources of uncertainty are: uncertainty arising from the limited information regarding the model parameters, and uncertainty arising from the future conditions (because we are planning for the future). The first source of uncertainty leads to stochastic problems while the second source leads to the dynamic problems. In the 'static stochastic location problems' the fundamental assumption is that we do not have all information on the parameters values, although we can estimate them, usually with a probability. The second fundamental assumption is that the parameters do not change over the time, which implies they are 'static' rather than 'dynamic'. When dealing with stochastic location problems two approaches are stochastic planning and scenario-based planning. In these approaches, the goal is to robustly locate facilities locations such that they have good performance (given some criteria) under different cases with known probabilities.

In the dynamic models, the assumption is that the parameters are known, but they are not static. That said they may change over the planning horizon. In the worst case, these changes might lead to replanting and re-location of the facilities (here, ambulances). Under this condition, the optimization is to make a trade-off between re-optimization and/or the current planning.

The remainder of this paper is organized as follows. First, we will look at the previous studies in the context of the ambulance location problem including the modelling and solution strategies. Then, we propose two mathematical models to locate ambulances under stochastic demands. The first model, does not allow for ambulance movement between different locations while the second model allow for this. We illustrate the models by using several examples. Finally, we present the computational results and show how the models would lead to improvements in the planning. The paper ends with the conclusion.

# 2. Literature Review

ReVelle (1989) have considered a probabilistic maximal covering location problem which they call it maximum available location problem (MALP). The objective is to locate p servers with the objective of maximizing the coverage of a zone (demand) within a time limit with a reliability factor (thus, probabilistic maximal covering location). They modelled the problem as an integer programming model and solved it for instances of the transportation network of Baltimore city. Gendreau (2006) studied the Maximal Expected Coverage Relocation Problem where the goal is to provide a dynamic relocation strategy for emergency vehicle waiting sites in such a way that the expected covered demand is maximized and the number of waiting site relocations is controlled. They formulated the problem as an integer linear program and applied it to emergency medical services data from the Montreal area. Nozari et al (2015) located vehicles in order to cover potential future demand effectively. They developed a multi-period version, taking into account time-varying coverage areas, where vehicles are allowed to be repositioned in order to maintain a certain coverage standard throughout the planning horizon. They formulated it as a mixed integer program optimizing coverage at various points in time simultaneously, and developed a variable neighborhood search. Simultaneously making location and dispatching decisions can potentially improve some performance measures. Sadeghi et al (2021) developed a mathematical formulation which considers location and dispatching decisions while congestion is modelled as well. They solved the model by using a Genetic Algorithm.

Aboueljinane (2013) studied computer simulation models used for the analysis and improvement of EMS. They reviewed existing research on simulation models and the associated results. The models are simulation, mathematical programming and queuing theory models. Berman (2013) studied the maximum covering location problem on a network with travel time uncertainty represented by different travel time scenarios. Three models are studied: expected covering, robust covering, and expected robust covering. Exact and approximate algorithms are developed. The results of applying the models to the analysis of the location of fire stations in the city of Toronto show that the current system design is quite far from optimality. McLay and Mayorga (2013) formulate a model for determining how to optimally dispatch servers to prioritized customers given that dispatchers make classification errors in assessing the true customer priorities. Their model determines how to optimally dispatch ambulances to patients to maximize the expected coverage of true high-risk patients. Naoum-Sawaya & Elhedhli (2013) presented a two-stage stochastic optimization model for the ambulance redeployment problem that minimizes the number of relocations over a planning horizon while maintaining an acceptable service level. They applied the model to the real-world dataset of the Region of Waterloo Emergency Medical Services.

**2.1. A quick review on the history and concepts of emergency medical services**

American College of Emergency Physicians (ACEP) defines the Emergency Medical Services (EMS) as the all public services attaching to the people lives. These services include medical and health services for those affected, for example, in accidents or under any other medical problem, for example, stroke that require such services urgently (thus, a life threatening conditions). Obviously, such a system requires operations and cooperation of many elements, including people who are at the incident and call for emergency service and perform some preliminary care until the ambulance arrives, and those practitioners who provide the medical treatments for the patients (Nozari & Szmelter ,2018).

The history of the EMS shows that the 60's has been a revolutionary for these services. The advancement in the medical science, in particular, under emergency conditions and expansion of cities, have made the need for such services more crucial. In 1966, the EMS has established in the US highways and major roads. In 1967, the EMS established for patients with heart diseases, in Ireland. The first program with specialization in EMS was established at the Cincinnati University in the US, in 1972. In 1973, the EMS rules and regulations were developed in the US. According to this regulation, such an EMS must include the following elements: Staffs, medicine and equipment, appropriate educational system, information and dispatching systems, transportation, hospitals and medical centers, special care units (at hospitals), public access to the service, data recoding standards as well as public awareness. In 1984, the EMS program for children established in the US.

In Iran, after an accident at the Mehrabad International Airport in 1976, the lack of existence of such a system was quite obvious. The destruction of the ceiling of the main lounge of the airport, which was caused because of the turbulences of the big airliners and weak structural building, left 16 dead and 11 wounded. The lack of an EMS caused this accident to be considered as one of the worst accidents in Iran, where no service was in place to transport injuries to hospitals and medical centers. This was to establish Tehran Emergency Information Centre, called 115 (this number is reserved in Iran to dial for such services), in 1976. When established, the Iran's EMS was one of the advanced systems in the world.

The EMS is divided into two groups: Anglo-American and Franco-German. In the Franco-German the equipment is taken into the patient and the physician is located inside the ambulance, while in the Anglo-American system, the technicians who have been educated the first aids, take care of the patients and undergo the emergency procedures; then, they transfer the patients to hospitals. The Iran's EMS, when established, was the Anglo-American system.

Generally, the EMS is designed to provide the basic medical needs to patients under urgent conditions and with regard to the following three objectives: saving patient's life, avoiding more damage to the patient and quick recovery of the patients. Following this, these six steps are the major elements of an EMS, which are shown as a star as illustrated in Figure 1.
1. Diagnosis after receiving an emergency call;
2. Reporting the accident;
3. Quick response of the medical team;
4. Providing the appropriate and efficient treatment at the accident;
5. Monitoring the patient on the way to the hospital and/or the medical Centre;
6. Transporting the patient to the hospital considering his/her medical conditions and needs.

Around the world, EMS is provided by either governmental or private agencies. Based on this, the following categories can be considered.

- Governmental EMS: Ambulances and EMS staff are mostly located next to police or fire stations. For instance, this is the case of Iran. In England, this kind of service is offered only in big cities. In some countries such as the US, France, Germany and Japan, among others, firefighters are the first aid who arrive at the accident. Here, ambulances are under the firefighter operation, and if needed, the ambulances are used to transport the patients to the hospitals.
- Voluntarily EMS: Voluntarily providing EMS and transporting patients to hospitals is another category of EMS. In many events, for instance, big sport events, the voluntarily aid groups are on site to provide the service. As a world level example, Red Cross organization provides such services voluntarily in many countries.
- Private EMS: In some countries, the private agencies provide the services, including the staff and the ambulances, under the umbrella of the governmental rules and regulations.
- Combined EMS: In this service, all EMS staffs are familiar not only with the medical emergency procedures but with those procedures under police and fire operations. This is an efficient way, in particular, in highly crowded places such as airports, or in remote and low density places where it is not economical or possible to have separate systems for EMS, police and fire. Thus, in all conditions, an experienced and flexible team can be at the accident location.

- Hospital-based EMS: The hospitals might provide such EMS to the geographical zones they cover. As the EMS is under coverage of the hospitals, they may provide different levels of service according to the facilities available at the hospitals.

Most services in EMS are those related to the following cases:
1. Heart failure: these services include saving patients' lives which can be fatal because of the heart failure. Immediate treatments have the highest priority here.
2. Trauma: Unlike heart services, here, the highest priority is to transfer the patients to the hospitals and medical Centers. Many health systems recommend only the first aids such as the keeping the patients stable, at the accidents.
3. Breathing disorder: in a stroke or when the patient has difficulties in breathing, the highest priority is to get to the patient in the quickest possible time to the hospital and medical Centre. There are several emergency procedures in such accidents.
4. Children: children from 5 to 10% of the total demand for the EMS. Their ages cause different services than adults. Most demands in this case are because of breathing disorders.

## 3.The Emergency Vehicles Types and Configurations

Apart from infrastructure and systems including communications facilities which are required for efficient operational of an EMS, vehicles and ambulances are an essential part of an EMS. In fact, these vehicles are equipped with the first aids, to ensure the patients are safely transported to a more equipped and advanced medical Centre such as a hospital. Typically, three types of ambulances are available.
1. Standard truck: in this ambulance the patients are separated from the driver. The ambulance is equipped with the Basic Life Support (BLS).
2. Standard van: in this ambulance, the patient room is not separated from the driver.
3. Van chassis with modular box: this ambulance is a type of van vehicle; however, the patient is separated from the driver. This type of ambulance includes the full monitoring systems and Advanced Life Support (ALS).

The ambulances must include sirens and flashing and rotating light bulbs. Along with the use of the star of life and the word 'ambulance' spelled in reverse is highly effective for drivers (and pedestrians) to recognize an ambulance. It is required that the drivers are experienced enough and are broadly familiar with the roads to get the patients to the hospitals in the quickest possible time. The drivers have the authority to break certain driving rules like the red traffic light and speed limit, to ensure the patient is transported to the hospital in the quickest time possible. Usually, the standard distance for using the ambulance is under 240 KM. However, under certain conditions such as traffic congestion, air ambulance might be preferred. Following these priorities, determining the best locations of ambulances has tremendous impact on patient lives. This is the major motivation for such rich research on this problem. However, the research on the problem under stochastic demand and travel times has not been well studied.

### 3.1. Problem statement and formulation

In this study, two mathematical models have been developed to allocate the ambulances to the stations to cover the arising demand at the demand zones. Both models take into account the different time lags of the planning horizon. This is very important as it helps to model the real-world with higher accuracy following the fact that the traffic changes over different times of the day. Moreover, both models consider the capacity of stations and the cost of using the ambulances. In the second model, however, the ambulances can be moved between stations; this might be beneficial as it may lead to cost reduction. Both models assume that the all the demands must be met.

## 3.2. Model I

The first model is based on the allocation problem. However, it has been extended to consider the different times of the day by splitting the day into several time slots. Moreover, an inventory has been added such that if there are ambulances at the end of the time slot, this can be used in the next coming time slots. This is illustrated in Figure 1.

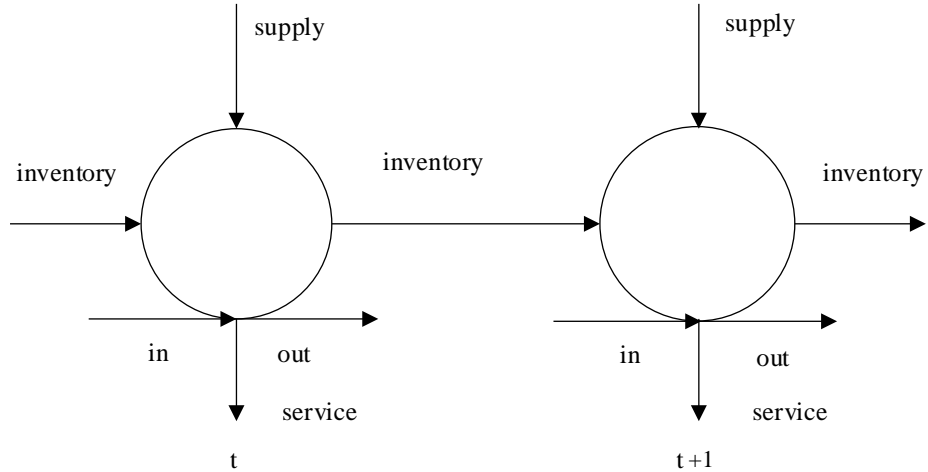

**Figure 1. First model is based on the allocation problem**

We start by explaining the notations and the decision variables.

Decision variables

$x_j^t$     The number of ambulances allocated to the station j at the time t

$s_j^t$     The number of ambulances dispatched from the station j at the time t

$I_j^t$     The inventory of ambulances at the station j at the end of the time t

$l_j^t$     The shortage of ambulance at the station j at the time t

Notations

$a_{ij}$     1 if the demand of the zone i can be covered by the station j, and 0 otherwise

$m$     The total number of the ambulances available in the system

$n_j^t$     The capacity of the station j at the time t

$c_j^t$     The fixed cost of keeping the ambulances at the station j at the time t

$c_j'^t$     The cost of dispatching the ambulances from the station j at the time t

$d_j^t$     The demand of the zone i at the time t

### 3.3. Constraints

The constraints of the Model 1 are vehicle inventory (we have enough number of vehicles), vehicle availability, covering the demand and capacity of the EMS stations. We shall explain the constraints in more details.

Vehicle inventory constraints: These constraints ensure that the vehicles can be used in the next time slots.

$$I_j^{t-1} + x_j^t - s_j^t = I_j^t \qquad \forall j,t$$

Vehicle availability: These constraints are to ensure that we have a limited number of available vehicles.

$$\sum_j x_j^t \leq m \qquad \forall t$$

Demand covering constraints: These are to ensure that the demands of all zones are to be covered:

$$\sum_j x_j^t a_j^t \geq d_i^t \qquad \forall i,t$$

$$\sum_j s_j^t a_{ji} + l^t \geq d_i^t \qquad \forall i,t$$

$$\sum_j s_j^t + l^t = \sum_i d_i^t \qquad \forall t$$

$$s_j^t \leq x_j^t \qquad \forall j,t$$

Capacity constraints: These constraints are to ensure there is a capacity for every EMS station.
Objective function: The objective function of the Model 1 minimizes two costs: the cost associated with using the vehicles and the shortage in the number of vehicles. Here, we assume that the cost of using the ambulances is only dependent on the time of the day (different time slots) and the EMS stations.

$$min \sum_j \sum_t x_j^t c_j^t + l^t M + c_j'^t$$

### 3.4. Model 2

This model is based on the transportation problem. We extended the transportation problem such that the movement of vehicles between different stations are permitted. On the other hand, there is a transportation problem in every time slot, however, the source nodes are connected by edges to model the movement of vehicles between the stations. The concept of this model has been illustrated in Figure 2.

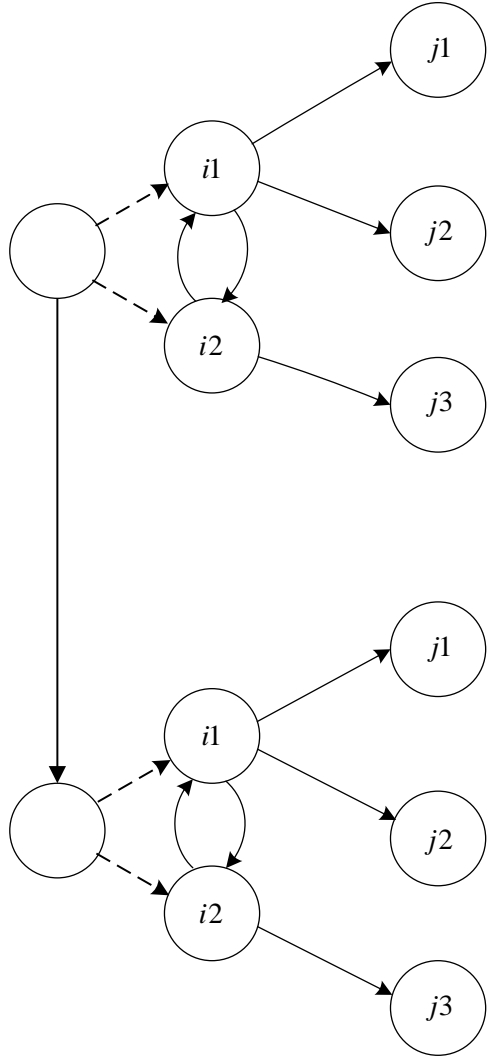

**Figure 2. Second model based on the transportation problem**

The first and the second constraints ensure that the vehicles can be moved from one-time slot to another. The third and the fourth constraints are the transportation problems constraints which model the capacity of the stations and the demand, respectively.

$$\sum_j \sum_i y_{ji}^t \leq I^0 + Inv^{t-1} \qquad \forall t$$

$$Inv^t = Inv^{t-1} + I^0 - \sum_i d_i^t \qquad \forall t$$

$$\sum_i y_{ji}^t a_{ji} + x_{j+}^t - x_{j-}^t \leq n_j^t \qquad \forall j,t$$

$$\sum_j y_{ji}^t a_{ji} = d_i^t \qquad \forall i,t$$

The objective function of the Model 2 is:

$$\sum_j \sum_t \sum y_{ji}^t c_j^t + l^t . M \qquad \forall i,j,t$$

The computational results

To validate the Model 2, we solve five instances in which we have considered the demand for vehicles arising randomly following a discrete uniform statistical distribution. All pre-processing have been implemented in programming language Python; the mathematical models have been coded in C++ and solved by CPLEX.

The parameters of these instances have been brought in Table 1.

Table 1. Parameters of the instances

| Parameters | Instance 1 | Instance 2 | Instance 3 | Instance 4 | Instance 5 |
|---|---|---|---|---|---|
| Number of vehicles | 100 | 200 | 200 | 200 | 200 |
| Number of time slots | 4 | 4 | 4 | 12 | 24 |
| Number of stations | 10 | 15 | 20 | 20 | 20 |
| Number of zones | 20 | 30 | 40 | 40 | 60 |
| LB on station capacity | 1 | 1 | 1 | 1 | 1 |
| UB on station capacity | 6 | 6 | 6 | 6 | 6 |
| LB on vehicle fixed cost | 6 | 6 | 6 | 6 | 6 |
| UB on vehicle fixed cost | 10 | 10 | 10 | 10 | 10 |
| LB on vehicle service cost | 2 | 2 | 2 | 2 | 2 |
| UB on vehicle service cost | 6 | 6 | 6 | 6 | 6 |
| LB on demand | 0 | 0 | 0 | 0 | 0 |
| UB on demand | 5 | 5 | 5 | 5 | 5 |

LB refers to lower bound and UB refers to upper bound. The details of the computational results have been brought in tables 2-6.

Table 2. Computational results (Example 1)

| Time slot | Number of vehicles in zones | | | | | | | | | | Shortage |
|---|---|---|---|---|---|---|---|---|---|---|---|
|  | 1 | 2 | 3 | 4 | 5 | 6 | 7 | 8 | 9 | 10 |  |
| 1 | 3 | 4 | 3 | 5 | 2 | 3 | 3 | 5 | 2 | 2 | 6 |
| 2 | 5 | 3 | 5 | 5 | 2 | 4 | 4 | 2 | 5 | 3 | 0 |
| 3 | 2 | 2 | 1 | 5 | 1 | 2 | 5 | 3 | 2 | 5 | 11 |
| 4 | 2 | 3 | 5 | 2 | 5 | 1 | 3 | 5 | 3 | 4 | 15 |

Table 3. Computational results (Example 2)

| Time slot | Number of vehicles in zones | | | | | | | | | | | | | | | Shortage |
|---|---|---|---|---|---|---|---|---|---|---|---|---|---|---|---|---|
|  | 1 | 2 | 3 | 4 | 5 | 6 | 7 | 8 | 9 | 10 | 11 | 12 | 13 | 14 | 15 |  |
| 1 | 3 | 5 | 3 | 3 | 5 | 1 | 1 | 4 | 5 | 3 | 2 | 1 | 3 | 2 | 3 | 17 |
| 2 | 1 | 1 | 5 | 1 | 5 | 4 | 1 | 3 | 5 | 3 | 4 | 4 | 5 | 4 | 2 | 22 |
| 3 | 3 | 3 | 4 | 5 | 1 | 2 | 5 | 3 | 4 | 4 | 4 | 3 | 2 | 2 | 4 | 3 |
| 4 | 5 | 2 | 3 | 5 | 1 | 1 | 5 | 5 | 1 | 3 | 1 | 4 | 3 | 5 | 3 | 7 |

Table 4. Computational results (Example 3)

| Time slot | Number of vehicles in zones | Shortage |
|---|---|---|

21

|   | 1 | 2 | 3 | 4 | 5 | 6 | 7 | 8 | 9 | 10 | 11 | 12 | 13 | 14 | 15 | 16 | 17 | 18 | 19 | 20 |    |
|---|---|---|---|---|---|---|---|---|---|----|----|----|----|----|----|----|----|----|----|----|----|
| 1 | 2 | 1 | 1 | 4 | 3 | 2 | 5 | 5 | 4 | 3  | 2  | 3  | 1  | 5  | 2  | 3  | 3  | 4  | 2  | 4  | 15 |
| 2 | 2 | 2 | 5 | 5 | 5 | 5 | 3 | 4 | 5 | 1  | 3  | 4  | 4  | 3  | 1  | 2  | 5  | 3  | 4  | 5  | 16 |
| 3 | 1 | 2 | 3 | 3 | 1 | 1 | 1 | 3 | 1 | 4  | 3  | 1  | 4  | 3  | 4  | 3  | 5  | 4  | 2  | 4  | 28 |
| 4 | 2 | 5 | 2 | 3 | 5 | 2 | 4 | 1 | 4 | 4  | 4  | 5  | 1  | 2  | 4  | 3  | 4  | 2  | 3  | 4  | 29 |

**Table 5. Computational results (Example 4)**

| Time slot | Number of vehicles in zones |  |  |  |  |  |  |  |  |  |  |  |  |  |  |  |  |  |  |  | Shortage |
|---|---|---|---|---|---|---|---|---|---|---|---|---|---|---|---|---|---|---|---|---|---|
|   | 1 | 2 | 3 | 4 | 5 | 6 | 7 | 8 | 9 | 10 | 11 | 12 | 13 | 14 | 15 | 16 | 17 | 18 | 19 | 20 |    |
| 1  | 2 | 5 | 3 | 4 | 5 | 1 | 5 | 4 | 3 | 2 | 2 | 2 | 1 | 2 | 1 | 3 | 1 | 3 | 2 | 3 | 21 |
| 2  | 5 | 1 | 4 | 3 | 2 | 4 | 3 | 2 | 4 | 3 | 3 | 5 | 2 | 1 | 2 | 2 | 2 | 2 | 5 | 2 | 32 |
| 3  | 1 | 3 | 3 | 1 | 3 | 5 | 2 | 3 | 3 | 4 | 2 | 2 | 1 | 4 | 3 | 3 | 3 | 1 | 4 | 5 | 16 |
| 4  | 1 | 1 | 3 | 4 | 5 | 1 | 4 | 3 | 3 | 5 | 4 | 1 | 1 | 4 | 4 | 4 | 5 | 4 | 1 | 1 | 33 |
| 5  | 3 | 1 | 2 | 4 | 5 | 2 | 2 | 4 | 4 | 1 | 1 | 4 | 4 | 4 | 4 | 1 | 1 | 1 | 2 | 5 | 16 |
| 6  | 5 | 5 | 2 | 2 | 1 | 1 | 5 | 1 | 2 | 1 | 1 | 2 | 4 | 3 | 4 | 4 | 5 | 1 | 1 | 4 | 23 |
| 7  | 1 | 2 | 2 | 3 | 3 | 2 | 2 | 3 | 2 | 5 | 1 | 3 | 1 | 3 | 2 | 1 | 1 | 1 | 2 | 3 | 42 |
| 8  | 4 | 5 | 3 | 5 | 5 | 2 | 1 | 4 | 2 | 4 | 4 | 4 | 4 | 2 | 3 | 4 | 1 | 1 | 3 | 5 | 3  |
| 9  | 1 | 2 | 4 | 3 | 4 | 3 | 2 | 5 | 2 | 1 | 2 | 5 | 2 | 4 | 5 | 3 | 4 | 3 | 1 | 5 | 27 |
| 10 | 2 | 3 | 5 | 5 | 5 | 5 | 4 | 5 | 4 | 1 | 4 | 1 | 4 | 1 | 2 | 4 | 4 | 1 | 1 | 2 | 14 |
| 11 | 3 | 3 | 4 | 2 | 5 | 4 | 2 | 4 | 1 | 4 | 5 | 5 | 2 | 4 | 3 | 5 | 5 | 4 | 2 | 4 | 4  |
| 12 | 3 | 5 | 5 | 4 | 4 | 1 | 2 | 3 | 2 | 5 | 3 | 3 | 3 | 2 | 5 | 4 | 4 | 5 | 2 | 1 | 23 |

**Table 6. Computational results (Example 5)**

| Time slot | Number of vehicles in zones |  |  |  |  |  |  |  |  |  |  |  |  |  |  |  |  |  |  |  | Shortage |
|---|---|---|---|---|---|---|---|---|---|---|---|---|---|---|---|---|---|---|---|---|---|
|   | 1 | 2 | 3 | 4 | 5 | 6 | 7 | 8 | 9 | 10 | 11 | 12 | 13 | 14 | 15 | 16 | 17 | 18 | 19 | 20 |    |
| 1  | 1 | 5 | 1 | 2 | 1 | 3 | 3 | 3 | 3 | 5 | 1 | 1 | 2 | 4 | 2 | 3 | 4 | 5 | 3 | 2 | 50 |
| 2  | 2 | 4 | 1 | 2 | 3 | 1 | 4 | 2 | 2 | 1 | 1 | 5 | 2 | 5 | 5 | 3 | 4 | 3 | 2 | 2 | 53 |
| 3  | 5 | 2 | 2 | 2 | 2 | 5 | 5 | 2 | 3 | 3 | 3 | 5 | 3 | 1 | 1 | 3 | 4 | 4 | 4 | 2 | 61 |
| 4  | 4 | 3 | 1 | 2 | 4 | 3 | 4 | 5 | 5 | 1 | 3 | 3 | 1 | 3 | 4 | 3 | 3 | 4 | 1 | 1 | 54 |
| 5  | 3 | 4 | 2 | 4 | 2 | 5 | 2 | 1 | 5 | 3 | 4 | 5 | 3 | 2 | 1 | 2 | 5 | 3 | 3 | 3 | 74 |
| 6  | 2 | 2 | 1 | 4 | 1 | 2 | 2 | 1 | 4 | 1 | 4 | 1 | 3 | 5 | 3 | 5 | 3 | 3 | 5 | 2 | 57 |
| 7  | 4 | 4 | 2 | 4 | 4 | 4 | 5 | 5 | 3 | 3 | 2 | 3 | 2 | 2 | 4 | 5 | 1 | 3 | 1 | 2 | 68 |
| 8  | 1 | 3 | 3 | 4 | 3 | 2 | 1 | 2 | 3 | 1 | 2 | 2 | 5 | 4 | 4 | 4 | 3 | 1 | 4 | 5 | 34 |
| 9  | 3 | 5 | 4 | 4 | 5 | 5 | 2 | 5 | 5 | 1 | 5 | 4 | 2 | 5 | 2 | 2 | 2 | 3 | 3 | 3 | 66 |
| 10 | 5 | 4 | 4 | 2 | 1 | 3 | 3 | 1 | 4 | 3 | 2 | 1 | 1 | 2 | 1 | 2 | 3 | 2 | 2 | 2 | 83 |
| 11 | 5 | 4 | 5 | 1 | 4 | 2 | 5 | 5 | 3 | 3 | 3 | 1 | 2 | 1 | 2 | 3 | 1 | 4 | 4 | 3 | 65 |
| 12 | 4 | 3 | 2 | 2 | 1 | 2 | 2 | 3 | 5 | 5 | 3 | 5 | 2 | 1 | 5 | 3 | 3 | 3 | 4 | 4 | 63 |
| 13 | 3 | 1 | 3 | 2 | 3 | 5 | 1 | 3 | 3 | 3 | 5 | 1 | 1 | 2 | 5 | 2 | 1 | 3 | 4 | 2 | 37 |
| 14 | 1 | 2 | 2 | 2 | 1 | 3 | 2 | 4 | 5 | 3 | 5 | 2 | 3 | 5 | 1 | 4 | 5 | 2 | 3 | 4 | 50 |
| 15 | 2 | 2 | 2 | 1 | 5 | 2 | 3 | 1 | 2 | 1 | 1 | 2 | 5 | 3 | 5 | 5 | 3 | 4 | 5 | 1 | 70 |
| 16 | 4 | 1 | 2 | 5 | 2 | 3 | 1 | 4 | 4 | 1 | 5 | 5 | 4 | 5 | 4 | 5 | 1 | 5 | 5 | 3 | 64 |
| 17 | 5 | 1 | 2 | 1 | 1 | 2 | 2 | 4 | 1 | 4 | 4 | 3 | 2 | 2 | 1 | 5 | 4 | 5 | 5 | 3 | 71 |
| 18 | 3 | 4 | 2 | 3 | 3 | 5 | 5 | 5 | 3 | 2 | 1 | 4 | 3 | 3 | 2 | 2 | 3 | 2 | 2 | 1 | 62 |
| 19 | 2 | 4 | 1 | 5 | 2 | 3 | 1 | 2 | 4 | 1 | 2 | 3 | 1 | 5 | 2 | 2 | 1 | 5 | 5 | 4 | 54 |
| 20 | 3 | 4 | 4 | 3 | 1 | 3 | 5 | 2 | 2 | 4 | 4 | 4 | 2 | 2 | 5 | 1 | 3 | 2 | 4 | 5 | 41 |
| 21 | 1 | 1 | 2 | 1 | 3 | 3 | 5 | 3 | 4 | 1 | 3 | 2 | 4 | 5 | 1 | 1 | 4 | 3 | 4 | 2 | 56 |
| 22 | 2 | 3 | 4 | 3 | 4 | 2 | 3 | 4 | 3 | 3 | 4 | 5 | 2 | 1 | 2 | 4 | 2 | 2 | 3 | 1 | 57 |
| 23 | 5 | 1 | 4 | 5 | 5 | 5 | 1 | 5 | 4 | 4 | 5 | 2 | 3 | 2 | 4 | 2 | 2 | 4 | 3 | 2 | 73 |
| 24 | 4 | 1 | 4 | 1 | 4 | 4 | 1 | 4 | 1 | 2 | 2 | 2 | 2 | 5 | 1 | 5 | 5 | 1 | 5 | 1 | 43 |

# 4. Conclusion

In this paper, two mathematical programming formulations to allocate vehicles to stations such that the demand of all zones is covered, have been developed. In both models, the concept of time slots has been introduced to better model the real-world scenarios. Hence, a planning day is split into several time slots which may have different operational conditions. The model 1 which is based on the location problem allocates the ambulances, in each time slot, to stations such that all demand is covered and the associated costs are minimized. In the Model 2, the movement of ambulances between stations have been introduced.